\documentclass[preprint,prb,aps,titlepage,byrevtex,floatfix]{revtex4}
\usepackage{epsfig}
\usepackage{rotating}
\begin{document}
\title{Efficient computation of free energy of crystal phases
       due to external potentials by error-biased Bennett acceptance ratio 
       method}
\author{Pankaj A. Apte}
\affiliation{Department of Chemical Engineering,
         Indian Institute of Technology Kanpur,
         Kanpur,
         U.P, India  208016}
\date{\today}
\begin{abstract}
Free energy of crystal phases is commonly evaluated by thermodynamic integration (TDI)
along a reversible path that involves an external potential.  A persistent problem
in this method is that a significant hysteresis is observed due to differences in the
center of mass position of the crystal phase in the presence and absence of the external 
potential. To alleviate this hysteresis, a constraint on the translational degrees of freedom
of the crystal phase is imposed along the path and subsequently a correction term
is added to the free energy to account for such a constraint.  In this work, we propose
a new methodology termed as error-biased Bennett Acceptance ratio (EBAR) method
that effectively solves this problem without the need
to impose any constraint.  This method is simple to implement 
and it does not require any modification to the path.
We show the applicability of this method in 
the computation of crystal-melt interfacial energy by cleaving 
wall method [J. Chem. Phys., {\bf 118}, 7651 (2003)] 
and bulk crystal-melt free energy difference by constrained fluid $\lambda$-integration
method [J. Chem. Phys., {\bf 120}, 2122 (2004)] 
for a model potential of silicon.  
\end{abstract}
\maketitle
\section{Introduction}
Accurate evaluation of free energy of a crystal phase by simulation is important
to predict many important thermodynamic properties including
the melting temperature,~\cite{FRENKEL02} the relative stability of 
different crystal phases~\cite{FERNANDEZ04} (i.e., polymorphism), and 
the interfacial free 
energy of crystal phases.~\cite{DAVIDCHACK03,GROCHOLA05}
A common method of free energy evaluation is thermodynamic integration
(TDI), in which an external potential is imposed on the crystal phase to ensure
thermodynamic reversibility,~\cite{FRENKEL02}
and the method requires that the resulting free energy change be computed accurately.  
However, the external potential causes center of mass (CM) position of the crystal phase
to change relative to that in the absence of the external potential and
this leads to hysteresis in the free energy computation.
This problem has been reported to occur in many TDI methods including 
Einstein crystal method,~\cite{BROUGHTON83,FRENKEL02} 
constrained fluid $\lambda$-integration method,~\cite{GROCHOLA04,APTE06-1} 
surface free energy calculation of crystal phases,~\cite{GROCHOLA05}
and direct computation of crystal-melt interfacial energy by cleaving wall method.~\cite{DAVIDCHACK03}  

A variety of constraints are imposed on the translational degrees of freedom of the crystal phase 
to address this problem.  In Einstein crystal method, CM is fixed during 
thermodynamic integration along the path.~\cite{FRENKEL02}  
In direct computation of crystal-melt free energy difference, CM
is either fixed~\cite{GROCHOLA04} or confined to a small region.~\cite{APTE06-1}  A fixed particle 
constraint was also proposed for this method.~\cite{APTE06-1}  In surface free energy calculation of
Au[110] crystal phase,~\cite{GROCHOLA05} CM velocity was artificially controlled by the
(otherwise) dormant external potential during Molecular Dynamics (MD) simulation. In the computation
of crystal-melt interfacial energy $\gamma$, two crystal layers were immobilized by assigning them an infinite
mass throughout the four stage integration path.~\cite{DAVIDCHACK03} 
It can be seen from the above examples that although cause of the hysteresis is the same,  
every TDI path requires its own ad-hoc 
procedure to constrain the translational degrees of freedom of the crystal phase.
Further, the use of a constraint entails computation of the correction term to obtain free energy difference
between the unconstrained phases.  In case of Einstein crystal method,~\cite{FRENKEL02} the correction term
can be computed analytically and the effort required is minimal.   When CM is confined
to a small region in
constrained fluid $\lambda$-integration method, the correction term needs to be computed by
separate simulations.~\cite{APTE06-1}   In cleaving wall method,~\cite{DAVIDCHACK03}
$\gamma$ was computed for two different system sizes in order
to determine the effect of the constraint.  Thus, the estimation of the correction term 
is, in itself, computationally expensive in many cases.  

In this article, we propose a new methodology termed as error-biased  
Bennett acceptance ratio (EBAR) method 
(to be explained in  sec. II) to efficiently compute
the free energy change resulting from imposition of the external potential on the crystal phase,
{\it without} requiring the use of any constraint.  
We demonstrate two applications of this methodology in sec. III.  
We first compute the free energy of the crystal phase due to repulsive wall potential 
in cleaving wall method.~\cite{DAVIDCHACK03,APTE08}  
Next, we compute the 
free energy difference resulting from the imposition of 
an attractive Gaussian well potential on the
crystal phase in constrained fluid $\lambda$-integration method.~\cite{GROCHOLA04, APTE06-1}  
Both calculations are performed
for the Stillinger and Weber~\cite{STILLINGER85} potential of silicon.  
Finally, we summarize our results and discuss  
further possible applications of this methodology
in sec. IV. 
\section{Error-biased Bennett acceptance ratio method}

In this section, we first describe the Bennett acceptance ratio method~\cite{BENNETT76} 
(BAR) which is routinely
employed to compute the free energy difference between two states.~\cite{SHIRTS05,MU06,LU03}  
We then describe the error biased Bennett acceptance
ratio (EBAR) method in the context of the present problem.
According to the BAR method,~\cite{BENNETT76} Helmholtz free 
energy difference $\Delta F=F_1 -F_0$  between two equilibrium states `0' and `1' 
is given by the following equation:~\cite{BENNETT76}
\begin {equation}
 \Delta F =\log \frac {\sum_1 f(\phi_0-\phi_1+C)}
                    {\sum_0 f(\phi_1-\phi_0)-C)}
                       +C - \log \frac{n_1}{n_0} ,
\label{eq:BAR1}
\end {equation}
where $C$ is a constant, 
$f(x)=1/(1+e^x)$ is the fermi function, 
$\sum_0$ and $\sum_1$ represent the sums over fermi functions sampled in `0' 
and `1' ensembles, respectively.  The symbols $\phi_0$ and $\phi_1$ in
Eq.~(\ref{eq:BAR1}) represent the total configurational energies and 
$n_0$ and $n_1$ are the number of
sampled fermi functions in the above two ensembles.  Please note that we have
expressed $\Delta F$, $\phi_0$, $\phi_1$, and $C$ in units of $k_B T$ and we will
follow this convention throughout unless otherwise stated explicitly.
The error (in units of $k_B T$) in the free energy estimate is given by 
$\sigma^2=\langle(\Delta F - \Delta A)^2\rangle$, 
where $\Delta A$ is the expectation (correct) value 
of the free energy difference. 
When we are in the large sample regime (i.e., both $\sum_0$ and $\sum_1$
are reasonably accurate), the variance can be
approximated as~\cite{BENNETT76}
\begin {equation}
\sigma^2  \approx \frac{\langle f^2 \rangle_0 - \langle f \rangle_0^2}
                {n_0 \langle f \rangle_0^2} 
              +\frac{\langle f^2 \rangle_1 - \langle f \rangle_1^2}
                {n_1 \langle f \rangle_1^2}.
\label{eq:err1}
\end {equation}
where $\langle f \rangle_0 $ is the ensemble average of $f(\phi_1-\phi_0-C)$ 
evaluated in ensemble 0 and $\langle f \rangle_1$ is the ensemble average of 
$f(\phi_0-\phi_1+C)$ evaluated in ensemble 1.
Bennett showed that the value of $C=C_B$ 
which minimizes $\sigma^2$
is given by the following expression~\cite{BENNETT76}
\begin {equation}
\Delta F = C_B - \log \frac{n_1}{n_0}.
\label{eq:BAR2}
\end {equation}
This value of $C$ corresponds to the condition~\cite{BENNETT76} that $\partial \sigma^2/\partial C=0$.
Thus, in the BAR method, the optimum estimate of $\Delta F$ is obtained by solving Eqs.~(\ref{eq:BAR1}) 
and (\ref{eq:BAR2}) simultaneously.   The condition given by Eq.~(\ref{eq:BAR2}) can also be expressed
as $\sum_0 = \sum_1$.  The important requirement for the applicability of the BAR method is
that the large sample regime should be achieved in both the ensembles.

Considering the present problem, let's denote the crystal phase with and 
without an external potential as state 1
and 0, respectively.  In such a case, the instantaneous configurational energies 
in the two states are given by $\phi_0 $=$U$ and
$\phi_1 $=$ U + U_{ext}$, where $U$ is the potential
energy due to interactions between the particles and $U_{ext}$ is  
the external potential energy.  Because of the influence of $U_{ext}$, average
position of center of mass of the crystal phase will be different in the two states.
In order to sample the perturbation
energy $\phi_0 - \phi_1$  efficiently in state 1 the following two conditions are necessary.
(i) The important configurations must overlap to a large extent with those of state 0,
which is possible if the external potential does not affect the crystal structure significantly.  
(ii) Further, the external potential must be sufficiently strong so that  
the CM is localized at the average position ${\bf R_1}$.
If this latter condition is not satisfied, the number of important configurations (corresponding
to all possible CM positions) becomes
too large and cannot be sampled in a finite length simulation.
In state 0,  the important values of the perturbation energy $\phi_1 - \phi_0$
are those for which the CM position is ${\bf R}_1$ 
(corresponding to the state 1) 
and are therefore not likely to be sampled in a simulation of reasonable length 
because the CM position can fluctuate wildly in state 0 due to the absence of the
external potential.
This leads to a large error in the estimate of $\sum_0$ in Eq.~(\ref{eq:BAR1})
for a given value of $C$ and $n_0$.  

When conditions (i) and (ii) are satisfied, 
the value of $\sum_1$ in Eq.~(\ref{eq:BAR1}) can be estimated accurately 
in a simulation of reasonable length.  
However, due to poor estimate of $\sum_0$, 
the large sample regime is not achieved at $C=C_B$ and 
the BAR method 
fails.  To circumvent this difficulty, we propose the EBAR method, in which we 
choose an appropriate value of $C=C_E$ (away from $C_B$)
such that the estimate of both $\sum_0$  and $\sum_1$ 
are reasonably accurate and hence the large sample regime is
achieved.   For this purpose, it is important to note the following two points: 

(a) When the value of $C$ is chosen such that $\sum_0 \ge 1$, 
$\log \sum_0 \approx \log \langle \sum_0 \rangle$,~\cite{BENNETT76} 
where $\langle \sum_0 \rangle$ is the expectation (accurate) value of $\sum_0$ at a given $C$ and $n_0$
(see Fig.~4 of Ref.~\onlinecite{BENNETT76} and the accompanying discussion).
In such a case, the main source of error will be due to $\sum_1$, which we expect to be
sufficiently accurate when the conditions (i) and (ii) given above are satisfied.

(b) In order to locate the exact value of $C$ where a large sample regime will occur, we
first consider two limiting cases.
When $C$ approaches $C_{+\infty} $, 
such that $x = (\phi_0-\phi_1+C) \rightarrow +\infty$, 
$f(x) \approx e^{-x}$ and $f(-x) \approx 1$.  In this case,
Eq.~(\ref{eq:BAR1}) reduces to the single stage free energy perturbation (FEP) formula 
$\Delta F =\log \langle \exp[-(\phi_0-\phi_1)] \rangle_1$,
where $\langle \cdots \rangle_1$ is the NVT ensemble average in state 1.  In this
limit $ {\partial \sigma^2}/{\partial C}=0 $, however, we shall be necessarily
in the small sample regime since we are not using any data from the state 0
in the estimation of $\Delta F$
[the FEP formula in state 1
can be considered as a limiting case~\cite{BENNETT76} of the acceptance ratio formula 
in Eq.~(\ref{eq:BAR1}) as $n_0 \rightarrow 0$].
Also at $C=C_B$, $\partial \sigma^2/\partial C = 0$ according to the BAR method, 
but the result corresponds to 
the small sample regime due to the poor estimate of $\sum_0$,
as discussed before.  For an intermediate value of $C=C_E$ ($ C_B < C_E < C_{+\infty}$),
the magnitude of $\partial \sigma^2/\partial C$ must attend a local maximum.  
The large sample regime will occur in the immediate vicinity 
of $C_E$, since as we move away from $C_E$ in either
direction ($C \rightarrow C_B$ or $C\rightarrow C_{+\infty}$), the magnitude of 
${\partial \sigma^2}/{\partial C}$ decreases to $0$ and we approach 
the small sample regime.  

To summarize the EBAR method, we choose $C=C_E$ such that
the magnitude of $|\partial \sigma^2/\partial C|$ is at its maximum
and $\sum_0 \ge 1$. 
It may be noted that a local maximum of 
$|\partial \sigma^2/\partial C|$ will also exist when $\sum_1 \ge 1$, 
but it will not correspond to the large sample regime due to poor estimate 
of $\sum_0$. 
In order that the EBAR method be successful, the
estimation of $\sum_1$ must be sufficiently accurate.  This is ensured by 
the conditions (i) and (ii) mentioned above. 
As in Ref.~\onlinecite{BENNETT76},
the above methodology can be extended straightforwardly
to isothermal isobaric ($NPT$) ensemble.
In this case, the values of the perturbation energies in above equations will be 
sampled during $NPT$ simulations
and Eqs.~(\ref{eq:BAR1})--(\ref{eq:BAR2}) yield 
the Gibbs free energy difference $\Delta G$ between states 0 and 1 
instead of the Hemholtz free energy difference  
$\Delta F$.
\section{Applications}
In what follows, we demonstrate two applications of the EBAR method for 
the Stillinger-Weber (SW)
potential~\cite{STILLINGER85} of silicon at the previously reported melting
point, i.e., at $T^*=0.0667$ ($1678$~K) and $P^*=0$~\cite{YOO04} (quantities with
the superscript * are dimensionless).  

\subsection{Cleaving wall method}

In cleaving wall method, bulk crystal and melt phases are combined reversibly to form an
interface and the work required for this change per unit interfacial area yields the crystal-melt interfacial
energy $\gamma$.~\cite{DAVIDCHACK03}  This technique is sufficiently precise so as to resolve the
anisotropy of $\gamma$ and has been applied to study anisotropy of interfacial energies
of hard-spheres,~\cite{DAVIDCHACK00} 
soft-sphere potential,~\cite{DAVIDCHACK05} and to model potentials of 
silicon~\cite{APTE08} and water.~\cite{HANDEL08} 
In the first stage of this process, crystal phase is cleaved at a predetermined
plane so that no particle can cross that plane.  The cleaving is done by introducing a repulsive wall 
consisting of one or more ideal crystal layers.  These layers are chosen to have the same orientation
as that of the interface being studied.  
This process is prone to hysteresis due to center of mass motion of the crystal phase 
since it changes the relative distance between the wall and the crystal layers.  
To prevent the hysteresis, Davidchack
and Laird~\cite{DAVIDCHACK03} immobilized particles in the two layers of the crystal phase sufficiently far from the
cleaving plane while applying the cleaving potential.  
To study the effect of this fixed layer constraint, 
thermodynamic integration was performed 
with the same interfacial area but with fewer crystal layers and the resulting value of $\gamma$ was found 
to be in agreement within error bars with that of the larger system,  which indicated that 
no correction term was necessary.~\cite{DAVIDCHACK03}  However, computation of $\gamma$ for two different
system sizes is computationally expensive since it involves implementation of all four stages of the cleaving
wall method.  We, therefore, explored the possibility that using EBAR method, we may be able to compute
free energy accurately without the need to immobilize particles in the crystal layers.

We performed the computation for (111) orientation of the silicon crystal phase, since for this orientation 
the layers on the
opposite sides of the cleaving plane are not symmetric.   As a result, when the CM of the crystal phase
fluctuates, it results in fluctuation of the cleaving potential, thus leading to hysteresis.  
For (100) orientation, this problem does not occur because of the symmetry of layers on the opposite
sides of the cleaving plane.  
As in Ref.~\onlinecite{APTE08}, we employed crystal phase
with 3024 particles and a simulation box with dimensions of $L_x=3\sqrt{2}a$, $L_y=3\sqrt{1.5}a$, and
$L_{z}=14\sqrt{3}a$, where $a=(8/\rho_C)^{1/3}$ is the unit cell length of the crystal phase
and $\rho_C = 0.452 \sigma^{-3}$ is the crystal density.   
The crystal phase contained 84 layers of (111) orientation in the z-direction, with
36 particles in each layer.  
The total configurational energy at a given value of $z$ is given by
\begin{equation}
    \phi({\bf r}^N;z) = U({\bf r}^N) + U_{ext} ({\bf r}^N;z),
\label{eq:dphi1}
\end{equation} 
where ${\bf r}^N = ({\bf r}_1, \cdots , {\bf r}_N)$ is the instantaneous configuration of the
particles, $U_{ext}$ is the cleaving potential exerted by the wall, and $z$ is the distance between
the wall and the cleaving plane.
The cleaving potential consisted of a repulsive two-body term 
and a 3-body term derived from Stillinger and Weber potential as explained in Ref.~\onlinecite{APTE08}.
The cleaving wall was constructed of two ideal crystal layers of (111) orientation with 36 particles 
in each layer.~\cite{APTE08}
The cleaving planes were located at two boundaries of the simulation box in z-direction. 
The Helmholtz free energy change for cleaving of the crystal phase 
is given by the following expression:~\cite{DAVIDCHACK03}
\begin{eqnarray}
\label{eq:df1}
    \Delta F &=& \int_{z_i}^{z_f} dz\frac{\partial{F}}{\partial z} \\ \nonumber
             &=& \int_{z_i}^{z_f} dz  \left\langle \frac {\partial{\phi}}{\partial z} \right\rangle, 
\end{eqnarray} 
where $\langle \cdots \rangle$ is canonical ensemble average at a particular value of $z$.  
Following Ref.~\onlinecite{APTE08}, 
we chose the initial and final positions of the cleaving walls as $z_i=1.80$ and $z_f=0.75$, respectively.
We performed canonical ensemble Monte Carlo (MC) simulations at various values of $z$ 
with $50000$ MC steps for
equilibration and $2 \times 10^5$ steps for production run.   
Each MC step consisted of 3024 trial displacement moves.
The maximum value of the attempted displacement during the trial moves was adjusted during
the equilibration period to have an acceptance ratio of nearly 50 \%.  
The integrand in Eq.~(\ref{eq:df1}) and
the perturbation energy 
[($\phi_1- \phi_0$) or ($\phi_0-\phi_1$) required in Eq.~(\ref{eq:BAR1}) ]
was sampled after every MC step.  

Figure~\ref{fig:f1} shows the negative of the integrand in Eq.~(5) 
per unit area (obtained without applying any constraint) as a function of the distance $z$.  
The results and the corresponding error bars reported in Fig.~\ref{fig:f1} 
and elsewhere represent the mean and the standard error 
of the estimates obtained from block averages during the production run.
The ordinate in the plot is the magnitude of the force between the wall and 
the particles, which increases with decreasing $z$ because of the repulsive nature of
the wall.  The integrand shows large hysteresis throughout the path in the absence of the any 
constraint.  However, when we apply the fixed layer constraint 
the hysteresis disappears as seen in the inset of Fig.~\ref{fig:f1}.
When performing simulations with the constraint,~\cite{DAVIDCHACK03} 
particles in two central layers (layers numbering
42 and 43) in the z-direction were fixed, i.e., trial moves attempting to displace the particles in these 
layers were rejected with certainty during MC simulation.   
A qualitative comparison of the plots shown in  Fig.~\ref{fig:f1}  confirms that
fluctuations in the CM leads to the hysteresis. 
Note that due to the effect of the
constraint on the free energy, the values of the integrand 
in the inset of Fig.~\ref{fig:f1} are significantly different from those in the main plot.
Due to this hysteresis, the BAR method is expected to fail and hence we have applied the
EBAR method as explained below.

In order to compute the free energy difference, we
consider state 0 as the crystal phase with a cleaving wall distances of 
$z$=$1.8$ which  is equal to the cut-off distance
of the external potential.  Thus, in state 0 no external potential
acts on the system.  As we decrease the value of $z$, the influence of the external
potential acting on the system increases.  
In Figs.~\ref{fig:f2} and \ref{fig:f3}, we report the relevant details of the EBAR and the 
BAR methods when using data from two simulations
at $z=0.75$ (state 1) and $1.8$ (state 0).  
As seen in Fig.~\ref{fig:f2}, the BAR result, which corresponds to the 
condition $\partial \sigma^2/\partial C = 0$ (or equivalently $\sum_0=\sum_1$),
is in the small sample regime,~\cite{BENNETT76} since  $\sum_0 = \sum_1 << 1$.  Also, from the
slope of the curve at $C=C_B$ in Fig.~\ref{fig:f3}, we find that $\partial \Delta F/\partial C = -1$ which 
again confirms that the BAR result is in the small sample regime.~\cite{BENNETT76}  
Thus the optimization scheme fails and the BAR method result is far off from 
the actual value, as seen in Fig.~{\ref{fig:f3}}.  
In the EBAR method, we choose a value of $C_E$ (see Fig.~\ref{fig:f2}) such that 
the magnitude of $\partial \sigma^2/\partial C$ is maximum and
$\sum_0 \ge 1$.  The corresponding value of $\Delta F$ 
(see Fig.~\ref{fig:f3}) 
compares well with the accurate result reported in Ref.~\onlinecite{APTE08} using
the BAR method.  
At a large positive value of $C > 220$, the value of $\Delta F/A$ 
computed using the acceptance ratio formula 
Eq.~(\ref{eq:BAR1}) approaches $0.0175~\mbox{J/m}^2$, 
which corresponds to 
the FEP formula in state 1.  
Thus, the EBAR method (see Fig.~\ref{fig:f3}),  
yields more accurate results compared to single stage FEP method.   
This is expected since the EBAR method utilizes data from both the ensembles
unlike the FEP method which relies on data from ensemble 1 alone.

Figure~\ref{fig:f4} compares the computational effort required to obtain 
$\Delta F/A$ using different methodologies (see also Table~\ref{tab:t1}).
When considering more than two simulations ($N_s > 2$), 
the EBAR method is applicable only to the
last interval and hence is combined with the BAR method to obtain total
free energy difference.  
As an example, when considering 4 simulations ($N_s = 4$)
at $z=0.75$, $0.8$, $0.85$, and $1.8$,
the result reported under the combined method (C) (see Table~1 and Fig.~\ref{fig:f4}) 
is a sum of BAR results
from $z=0.75$ to $0.85$ and the EBAR result from $z=0.85$ to $1.8$.  
Figure~\ref{fig:f4} shows that
the  EBAR method yields accurate value with just two simulations while the BAR 
method requires at least 9 simulations.  
It can also be seen that numerical integration using  Gaussian
Quadrature (GQ) technique using 3 simulations
is close to the correct value, however the size of the error bar is still relatively 
large compared to the EBAR result. 
This clearly indicates that GQ 
will require 4 or more simulations to achieve the desired accuracy.  

We also tested for the convergence of the EBAR method in Fig.~\ref{fig:f4}  by using
simulation data at intermediate $z$ values.  For example
for $N_s = 2$, $4$, and $6$, the EBAR method was applied to the intervals 
from 
$z=0.75$ to $1.8$,
$z=0.85$ to $1.8$,
and $z=1.0$ to $1.8$, respectively.
As the width of the interval reduces, the extent of the overlap between the configurations 
in state 1 and state 0 will increase.  However, the CM becomes less localized in state 1
due to reduced strength of the external potential at higher value of $z$
and hence the estimation of $\sum_1$
becomes less accurate for a given $n_1$ as discussed in Sec.~II.  
Note also that the size of the error bars in Fig.~\ref{fig:f1} is larger for $z > 0.75$ 
compared to that for $z=0.75$, indicating a larger hysteresis due to CM fluctuations. 
Since the accuracy
of the EBAR method depends on $\sum_1$, we find that it becomes less accurate 
(although marginally) for smaller intervals (see EBAR/Combined results in Table~\ref{tab:t1}
and Fig.~\ref{fig:f4}).

\subsection{Constrained fluid $\lambda$-integration}

Grochola~\cite{GROCHOLA04} introduced constrained fluid $\lambda$-integration
method to compute the free energy difference between the crystal and the melt phases.   
A major advantage of this method compared to traditional TDI methods~\cite{FRENKEL02} 
is that it {\it directly} 
calculates the free energy difference between the melt and the crystal phases by constructing
a reversible path between the two phases. Thus, there is no need to connect the crystal and the 
melt phases {\it separately}
to reference phases of known free energy
and this gives more flexibility in terms of designing
the reversible path.   Grochola's method has been applied to calculate
melting temperature of complex potentials for Sodium Chloride,~\cite{EIKE05} 
Benzene and trizole.~\cite{EIKE06}  
Further, the method was extended to isothermal isobaric ensemble,~\cite{APTE06-1} 
and to computation of  melting temperature
of binary mixtures~\cite{APTE05} and interfacial free energy of crystal phases.~\cite{GROCHOLA05}

Here we compute Gibbs free energy difference between crystal and melt phases of silicon
by $NPT$ version~\cite{APTE06-1} of the constrained fluid $\lambda-$ integration method.
The expressions for the potential energy for the 3 stages of the reversible path are as
given below:
\begin{equation}
    \phi_1(\lambda_1) = (1- \eta \lambda_1) U,
\label{eq:phi1}
\end{equation}
\begin{equation}
   \phi_2(\lambda_2) = (1-\eta) U + \lambda_2 U_{ext},
	\label{eq:phi2}
\end{equation}
and
\begin{equation}
 \phi_3(\lambda_3) = [(1-\eta)+\lambda_3 \eta] U + (1-\lambda_3) U_{ext},
\label{eq:phi3}
\end{equation}
where $U$ is the potential energy due to interactions between the system particles,  
$\eta$ is a parameter controlling 
the extent to which strength of interaction is reduced in 
the first stage.~\cite{GROCHOLA04,APTE06-1}
$\lambda_1$, $\lambda_2$, and $\lambda_3$ are the parameters 
characterizing the three stages. 
The Gaussian external potential imposed during the second and the third stage is given 
by~\cite{GROCHOLA04,APTE06-1}
$U_{ext} = \sum_{i} \sum_{k} a \exp(-b r_{ik}^2)$, where the summation with respect
to $i$ is taken over all the particles, $r_{ik}$ is 
the distance between the $i{\mbox{th}}$ particle and $k{\mbox{th}}$ well, and
the summation with respect to $k$ is taken over all Gaussian potential wells 
within a certain cutoff distance of the $i$th particle.   
A constraint on the maximum possible volume $V_m$ along the path is 
imposed~\cite{APTE06-1} to ensure thermodynamic reversibility.  The Gibbs free energy
change for the third stage of the path is given by:~\cite{APTE06-1}
\begin{equation}
    \Delta G_{3} =  \int_{0}^{1} d\lambda_3 \left\langle 
                \eta U - U_{ext} \right \rangle, 
\label{eq:dg1}
\end{equation}
where $\langle \cdots \rangle$ represents the isothermal--isobaric ensemble average at a given
value of $\lambda_3$.  Similar expressions
apply for $\Delta G_1$ and $\Delta G_2$.~\cite{APTE06-1}

The simulation at a given state point along the path was performed in 
$NPT$ ensemble with $N=1000$ particles confined to a cubic simulation box 
under periodic boundary conditions. 
At a given state point, we used $15000$ MC steps for
equilibration and $20000$ steps for production, except at the end of 3rd stage, where
$10^5$ MC steps were used for production run from $\lambda_3=0.94$ to $1$.  
In each MC step, we attempted, on average,
two volume change moves and 1000 particle displacement moves.  The size of the
attempted changes was adjusted during equilibration steps so as to achieve an 
acceptance ratio of about $50 \%$.  During production run, we sampled the integrand for
TDI method and the perturbation energy for BAR method after every MC step.   
The Gaussian well parameters were chosen  
to be $a=-1.892 \epsilon$ and $b=8.0 \sigma$ in accordance with the criteria mentioned
in Refs.~\onlinecite{APTE06-1,APTE05}. 
The parameter $\eta$ in Eqs.~(\ref{eq:phi1})--(\ref{eq:phi3}) was assigned a value of $0.9$
following earlier work.~\cite{GROCHOLA04,APTE06-1}
Further, we chose the maximum volume to be $V_m^*= N/0.4$ at $T^*=0.0667$
and $P^*=0$ so that $V_m$ does not affect the free energy of the crystal or the melt
phases.~\cite{APTE06-1}

We found no hysteresis for the first two stages of the path as in earlier studies.~\cite{GROCHOLA04}  
In the third stage, however, we found hysteresis between $\lambda_3=0.99$ to $1.0$ as
seen in the inset of Fig.~\ref{fig:f5}.   The reason for this hysteresis is that
as ($\lambda_3 \rightarrow 1$), the influence of the external potential
on the crystal phase becomes negligible as can be seen in expression of $\phi_3$ in 
Eq.~(\ref{eq:phi3}), which results in fluctuations of the CM position.  
Note that in computations performed by Grochola (see Fig.~6 of Ref.~\onlinecite{GROCHOLA04}),
no hysteresis was seen in the third stage, 
because zero CM velocity was maintained during MD simulations.
In order to compute the free energy difference in stage 3,
we denote state corresponding to $\lambda_3=1$  
as state 1, in which no external potential acts on the crystal phase.  
Note that the role of states 0 and 1 is reversed compared what 
we discussed earlier since the external potential is
acting in state 0 ($\lambda_3 < 1$) in the present case.
As $\lambda_3$ is
decreased, the influence of the external potential increases.  
We compared BAR, EBAR, GQ methods on the basis
of number of simulations ($N_s$) performed at various values of $\lambda_3$ between $0.9$ and $1$
to obtain a given result.  For $N_s > 2$, the EBAR result is applicable only to the last
interval and hence we combine it with the BAR result for the rest of the intervals as 
explained before.
(Also note that Eq.~(\ref{eq:BAR1}) will yield $\Delta G$
instead of $\Delta F$ since we are dealing with $NPT$ ensemble as mentioned at the end
of sec. II).  

Figure~\ref{fig:f6} and \ref{fig:f7} show the results for the BAR and the EBAR results 
using two simulations performed at 
$\lambda_3=0.94$ (state 0) and  $1$ (state 1).  
As seen in Fig.~\ref{fig:f6}, the BAR result 
is in the small sample regime since $\sum_0 = \sum_1 << 1$.  Also, we
find that $\partial \Delta G/\partial C = -1$ at $C=C_B$ from the slope of
the plot in Fig.~\ref{fig:f7}, which again indicates a small sample regime.~\cite{BENNETT76}  
As a result, the BAR result shows a large deviation from the actual value as seen
in Fig.~\ref{fig:f7}.  
On the other hand, the EBAR result (which corresponds 
to the maxima of $|\partial \sigma^2/\partial C|$ such that $\sum_1 \ge 1$)
is quite accurate as seen in Figs.~\ref{fig:f6} and \ref{fig:f7}. 
Note that as $C$ approaches a large negative value $C < -1170$, the $\Delta G$ 
value approaches $-33.8~k_B T$ which corresponds to 
the FEP formula in state 0.  Thus, the EBAR method yields more accurate
results compared to both the 
single stage FEP method and the BAR method.

In Fig.~\ref{fig:f8}, we have tested the convergence of the results
in the interval from $\lambda_3=0.9$ to $1$.
(Note that the figure shows the total Gibbs free energy difference 
$\Delta G_T$ for the entire path and all the
results reported in the figure
include a contribution of $60.75 \pm 3~k_B T$
from $\lambda_1 = 0$ to $\lambda_3=0.9$ computed by BAR method).  
As can be seen in the figure, BAR method 
requires 12 simulations to obtain acceptable accuracy. 
On the other hand, EBAR (combined) method
yields sufficiently accurate value of $\Delta G_T$ with 4 simulations.  
Using the converged result in Fig.~\ref{fig:f8}, we find that the contribution of
the CM hysteresis to the total error in $\Delta G_T$ is about $\pm 2~k_BT$. 
We also found the thermodynamic integration using Gaussian Quadrature (GQ) 
method  yields sufficiently accurate result with just 2 simulations 
in comparsion to 4 simulations required by the EBAR method.
In this case, GQ technique is effective because
the integrand changes smoothly (although rapidly) as $\lambda_3$ approaches $1$ 
and moreover GQ does not require evaluation of the integrand at $\lambda_3 = 1$ 
which is most prone to hysteresis.  

As for the convergence of the EBAR method, we note that for $N_s = 2$ 
(see Table~\ref{tab:t1} and Fig.~\ref{fig:f8}), the EBAR result (applied in 
the interval from $\lambda_3=0.9$ to $1$)
deviates significantly from the accurate value.  
This is because the external potential starts affecting the structure of the 
crystal phase significantly [see Eq.~(\ref{eq:phi3})] 
for $\lambda_3 \le 0.9$ (state 0)
and hence the value of $\sum_0$ becomes less
accurate, as discussed in Sec.~II.
As we increase $N_s$, the EBAR (combined) result converges 
rapidly as seen in Fig.~\ref{fig:f8}.
This is because even for $\lambda_3=0.99$ the CM position is sufficiently localized
and hence $\sum_0$ evaluation is accurate.  This can also be seen in the inset of
Fig.~\ref{fig:f5}, which shows that the CM hysteresis becomes appreciable only
for $\lambda_3 > 0.995$. 

Finally, based on the value of $\Delta G_T$,
we also computed the melting temperature (T$_m$) by integrating the following equation
at $P^*=0$:~\cite{BROUGHTON83}
\begin{equation}
   {k_BT^2}\left [ \frac{\partial(\Delta G_T/k_BT)}{\partial T} \right]_{P,N} 
  = (\langle U \rangle_L + P \langle V \rangle_L )
     - (\langle U \rangle_S + P \langle V \rangle_S),
\label{eq:gt}
\end{equation}
where $\langle \cdots \rangle_L$ and $\langle \cdots \rangle_S$ denote the
$NPT$ ensemble averages for the liquid and crystal phases at the specified
temperature.   We found the value of T$_m$ to be $1675\pm 5$ K, based on EBAR (combined)
result with $N_s = 4$.
This is in close agreement with the value of $1678$~K obtained in Ref.~\onlinecite{YOO04} 
by crystal-melt coexistence simulations for Si(100) interface and the
value of $1691 \pm 20$ K obtained in Ref.~\onlinecite{BROUGHTON87}.  

\section{Summary}
In this work, we have shown that EBAR method efficiently calculates the free energy
of the crystal phase due to an external potential without requiring
use of a constraint on the translational degrees of freedom.  In this 
method, we nullify the error incurred due to poor sampling of the 
perturbation energy in state 0 (crystal phase without
the external potential) by adjusting the value of the shift constant
$C$ [see Eq.~(\ref{eq:BAR1})] so that the error in estimated free energy difference 
is completely due to state 1 (the state in which an external potential acts on the crystal phase),
where we expect the sampling of the perturbation energy to be sufficiently accurate.  
We have applied this technique to cleaving wall method, in which the crystal phase is subjected to
a repulsive cleaving potential~\cite{DAVIDCHACK03} and confined 
fluid $\lambda$-integration method,~\cite{GROCHOLA04} in which the crystal
phase is acted upon by attractive Gaussian potential wells located at the ideal crystal lattice sites.
In both cases, we found that EBAR method yields accurate values with reasonable computational effort and
and offers considerably improvement over both 
the single stage FEP method (in state 1) and the BAR method.  Note that
unlike the FEP method, the EBAR method utilizes information from both the ensembles.

It must be stressed that the
EBAR method is applicable only when the
BAR result is in the small sample regime.  Thus, the domains of applicability of the two
methods are mutually exclusive.  
With regard to the convergence, we found that the EBAR method ceases to be accurate in the
following limits: (i) When the
external potential is too weak so that the CM position is not localized as seen in cleaving
wall method (see Fig.~\ref{fig:f4} and Table~\ref{tab:t1}) 
and (ii) when the external potential is too strong
so as to affect the crystal structure significantly as in the case of constrained fluid 
$\lambda$-integration
method (see Fig.~\ref{fig:f8} and Table~\ref{tab:t1}).
Both of these conditions increase the error 
the computation of $\sum_1$ (corresponding to 
the state in which external potential is acting).  
The EBAR method is simple to implement since it only requires that 
the perturbation energies
in the two states be sampled and does not depend upon the existence of a 
reversible path connecting the two states.

In the constrained fluid $\lambda$-integration method, thermodynamic integration by GQ technique 
is found to be effective since the CM hysteresis is confined 
to the end of the integration path (see Fig.~\ref{fig:f5}).
However, the GQ technique suffers from the inherent drawback of the TDI method in that 
it depends upon the existence of a reversible path between the two thermodynamic states.  
Also, testing the convergence of this technique is relatively
expensive since the abscissa values for higher number of integration points do not 
coincide 
with those corresponding to lower number of integration points.~\cite{PRESS92}
These problems prevent general applicability of the GQ technique,
as in the case of the cleaving wall method where the
CM hysteresis occurs throughout the path (see Fig.~\ref{fig:f1}).  

We expect that EBAR method will also 
be useful in computing bulk crystal phase free energy by Einstein crystal method~\cite{FRENKEL02} 
and in computing surface free energy of crystal phases~\cite{GROCHOLA05} where the CM hysteresis
occurs. It seems possible to generalize the EBAR method to other free energy computations.
Our initial calculations indicate that
the EBAR method can also be applied to the cleaving
of the crystal-melt interface (state 4 of the cleaving wall method), 
which is considered as a major challenge 
in the computation of the crystal-melt interfacial energy.~\cite{DAVIDCHACK03,MU06} 
Here the crystal-melt interface fluctuates in the absence of the external potential while it is
held fixed when external cleaving potential is present.~\cite{DAVIDCHACK03}  
It will be interesting to explore the applicability of EBAR method
to the computation of the chemical potential 
by particle insertion-deletion technique.~\cite{FRENKEL02}  
Here the perturbation energies due to the particle 
insertion steps are sufficiently accurate while those due to the
particle deletion steps are not sampled efficiently.  
This situation is similar to the problem considered in this article.

\begin{acknowledgments}
The author would like to thank Professors X. C. Zeng and I. Kusaka for helpful discussions.
This work was supported by the research initiation grant provided by the Indian Institute 
of Technology, Kanpur.  
\end{acknowledgments}

%
\clearpage
%
%
\begin{center}
\begin{table}
\begin{ruledtabular}
\caption{ \label{tab:t1}
    {The number of simulations ($N_s$) and the free energy calculations obtained by the BAR (B), EBAR (E) and
    the combined (C) results.  In the combined results ($N_s > 2$), 
    the EBAR method is applied for the last interval 
    while the BAR method is applied for the rest of the intervals.  
    The second and the third column 
    relate to the cleaving wall method, while the last two columns relate to the constrained fluid 
    $\lambda$-integration method. The data contained in this table is also plotted in Figs.~\ref{fig:f4}
     and \ref{fig:f8}. 
    }} 
\begin{tabular}{cllll}
    {$N_s$}  &  {$z$ values}       & $\Delta F/A$ (J/m$^2$) & {$\lambda$ values} & $\Delta G_T/k_B T$\\ \hline
$2$                   &  $0.75,~1.80$       &  $0.041 \pm 0.002 $ (E)   & $0.9,~1.0$     & $28.6 \pm 7.1 $ (E)   \\
                      &                     &  $0.36  \pm 0.15$  (B)     &                &                     \\ \hline
$4$                   &  $0.75,~0.80,~0.85$ &  $0.0455 \pm 0.005$ (C)    & $0.9,~0.92,~0.94$ & $12.4 \pm 4.6$ (C) \\
                      &  $1.80$             &                            & $1.0$             & $-576 \pm 193 $ (B) \\ \hline
$6$                   &  $0.75,~0.80,~0.85$ &  $0.0054 \pm 0.009$ (C)    & $0.9,~0.92,~0.94$ & $9.0 \pm 4.4 $ (C)\\
                      &  $0.90,~1.00,~1.80$ &  $0.088  \pm 0.024$ (B)    & $0.95,~0.96,~1.0$ & $-379 \pm 130 $ (B) \\ \hline
$9$                   &  $0.75,~0.80,~0.85$ &  $0.0439 \pm 0.004$ (B)    & $0.9,~0.92,~0.94$ & $5.5 \pm 4.4 $ (C) \\
                      &  $~0.90,~1.00,~1.10$ &                           & $0.95,~0.96,~0.97$ & $-87.9 \pm 34.4$ (B) \\
                      &  $~1.20,~1.50,~1.80$ &                           & $0.98,~0.99,~1.0$ &                 \\ 
\end{tabular}
\end{ruledtabular}
\end{table}
\end{center}
\clearpage
\begin{figure}
  \caption{\label{fig:f1}The variation of the integrand in Eq.~(5)
     as a function of $z$ 
     at $T^*=0.0667$, and $\rho_C=0.452 \sigma^{-3}$ 
     for cleaving of Si(111) crystal phase.   The inset
     shows the same plot with a fixed layer constraint.   In applying this constraint
     the particles in the middle layers (layers numbering 42 and 43) were immobilized.
     Note that due to the effect of the constraint on the free energy, the value of 
    integrand at a given $z$ is significantly larger 
    in the inset plot.  
    The error bars are seen when these are larger than the size of the symbols.
   }
\end{figure}
\begin{figure}
  \caption{
    \label{fig:f2} 
    The crieteria for choosing values of $C$ for the BAR and the EBAR methods.  
    The data is generated
    from simulations done at $z=1.8$ (state 0 with the external potential) and 
    $z=0.75$ (state 1 without the external potential). 
    The BAR result corresponds to the condition that $\sum_0 = \sum_1$ while the
    EBAR result corresponds to the maximum of $|\partial \sigma^2/\partial C|$ such 
   that $\sum_0$ is of order unity or higher.  A local maximum also exists
   when $\sum_1 \ge 1$, but is not seen in the figure because of its small magnitude.
   }
\end{figure}
\begin{figure}
  \caption{\label{fig:f3} 
  The value of Free energy difference per unit area (in units of $J/m^2$)
  obtained by Eq.~(\ref{eq:BAR1}) for the same set of data as that in Fig.~\ref{fig:f2}.  
  The inset shows more detailed comparison of the EBAR result with that of Ref.~\onlinecite{APTE08}
  (dashed line).
  From the slope of the plot, we get $\partial \Delta F/\partial C = -1$ at $C=C_B$ indicating
  a small sample regime for the BAR result.~\cite{BENNETT76}
  At the two end points of the graph ($C=-10$ and $C=220$), the value of $\Delta F$ approach those
  obtained from the FEP formulas in the two ensembles.
   } 
\end{figure}
\begin{figure}
  \caption{\label{fig:f4} 
    Free energy change per unit area (in J/m$^2$) resulting
    for cleaving of Si(111) crystal phase without any constraint.  The
    error bars are seen when these are larger than the size of the symbols.  
    The abscissa represents the number of simulations ($N_s$) performed 
    at different values of $z$ (see Table~\ref{tab:t1}). 
    The horizontal line represents the result obtained using the BAR method in Ref.~\onlinecite{APTE08}.
    The combined result (applicable for $N_s > 2$), represents a combination of the EBAR and the
    BAR methods as explained in the text.   
   }
\end{figure}
\begin{figure}
  \caption{\label{fig:f5} 
   The variation of the integrand in Eq.~(\ref{eq:dg1}) 
    as a function of $\lambda_3$ 
     at $P^*=0.0$, $T^*=0.0667$, and $N=1000$ for the forward and 
     the backward runs for stage 3.
    The inset shows the region of maximum hysteresis.
   }
\end{figure}
\begin{figure}
  \caption{
    \label{fig:f6} 
    The crieteria for choosing appropriate values of $C$ for the BAR and the EBAR methods.  
    The data is generated from 
    simulations performed at $\lambda_3=0.94$ (state 0) and 
    $\lambda_3=1.0$ (state 1).  The external Gaussian potential acts in state 0 while it is
   absent in state 1 according to Eq.~(\ref{eq:phi3}). The values of the Gibbs free energy
    difference $\Delta G$ between the two ensembles are plotted 
    as a function of $C$ in Fig.~{\ref{fig:f7}}.
    The BAR result corresponds to the condition that $\sum_0 = \sum_1$ while the
    EBAR result corresponds to the maximum of $|\partial \sigma^2/\partial C|$ such 
   that $\sum_1 \ge 1$.   
   }
\end{figure}
\begin{figure}
  \caption{\label{fig:f7} 
  The value of Gibbs free energy difference (in units of $k_B T$) between $\lambda_3=0.94$ and $1.0$ 
  states.  The result is 
  obtained by Eq.~(\ref{eq:BAR1}) for the same set of data as that in Fig.~\ref{fig:f6}.  Note that
  the slope of the curve is $-1$ at $C=C_B$ indicating a small sample regime for the BAR result.
  The inset shows more detailed comparison of the EBAR result with the accurate result (dashed line)
  obtained by inserting sevral intermediate steps between
  $\lambda_3=0.94$ and $1$.  At the two ends of the plot ($C=-1170$ and $C=0$) $\Delta G$ values approach 
  those corresponding to the FEP formulas in the two ensembles.
   }
\end{figure}
\begin{figure}
  \caption{\label{fig:f8} 
   {
    Total Gibbs Free energy difference (in units of $k_B T$) between the melt and the crystal phases 
    obtained without applying any constraint.  All of the
    results include a contribution of $60.75 \pm 3~k_B T$ from 
    $\lambda_1=0$ to $\lambda_3=0.9$ as mentioned in the text.  
    The abscissa represents the
    number of simulations performed at various values of $\lambda_3$ between $0.9$ and $1$.
    The horizontal line
    represents the converged value obtained by the BAR method with
    $N_s=12$ (see the inset). The size of the error bar for this converged result is 
    of the same magnitude as the combined result with $N_s = 9$.  
   }}
\end{figure}
\clearpage
\begin{figure}
  \begin{center}
    \epsfig{file=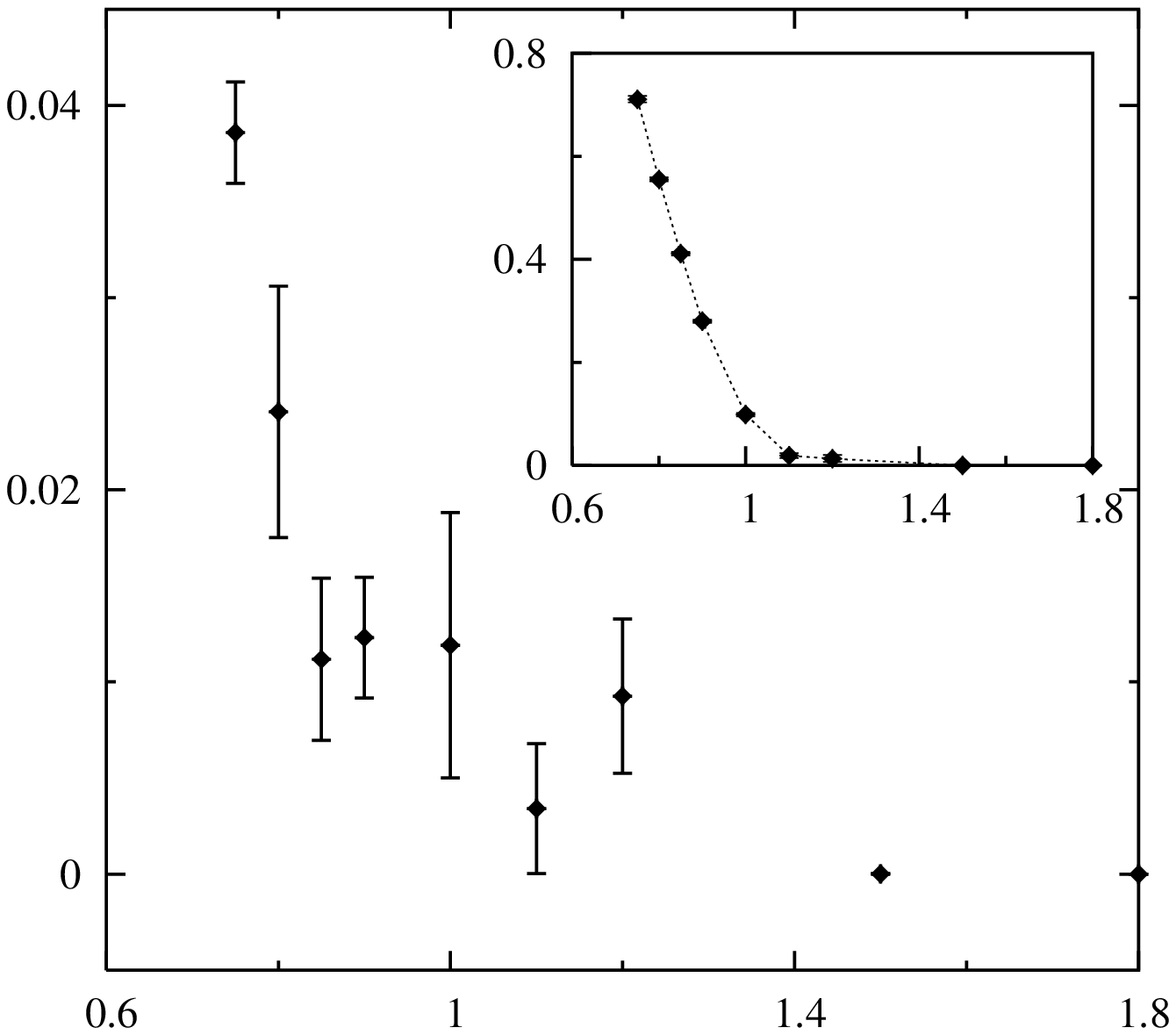,height=7.1cm,width=8.21cm}
    \put(-260,140){\large 
                $\frac{-1}{A}\frac{\partial F^*} {\partial z^*}$ 
                   }
    \put(-38,-7){$z/\sigma$}
    \put(-80,160){\footnotesize \mbox{fixed layer}}
    \put(-80,150){\footnotesize \mbox{constraint}}
    \put(-85,70){\footnotesize \mbox{No constraint}}
  \end{center}
\end{figure}
\clearpage
\pagenumbering{arabic}
\renewcommand{\thepage}{P. A. Apte, FIG.~\arabic{page}}
\begin{figure}
  \begin{center}
    \epsfig{file=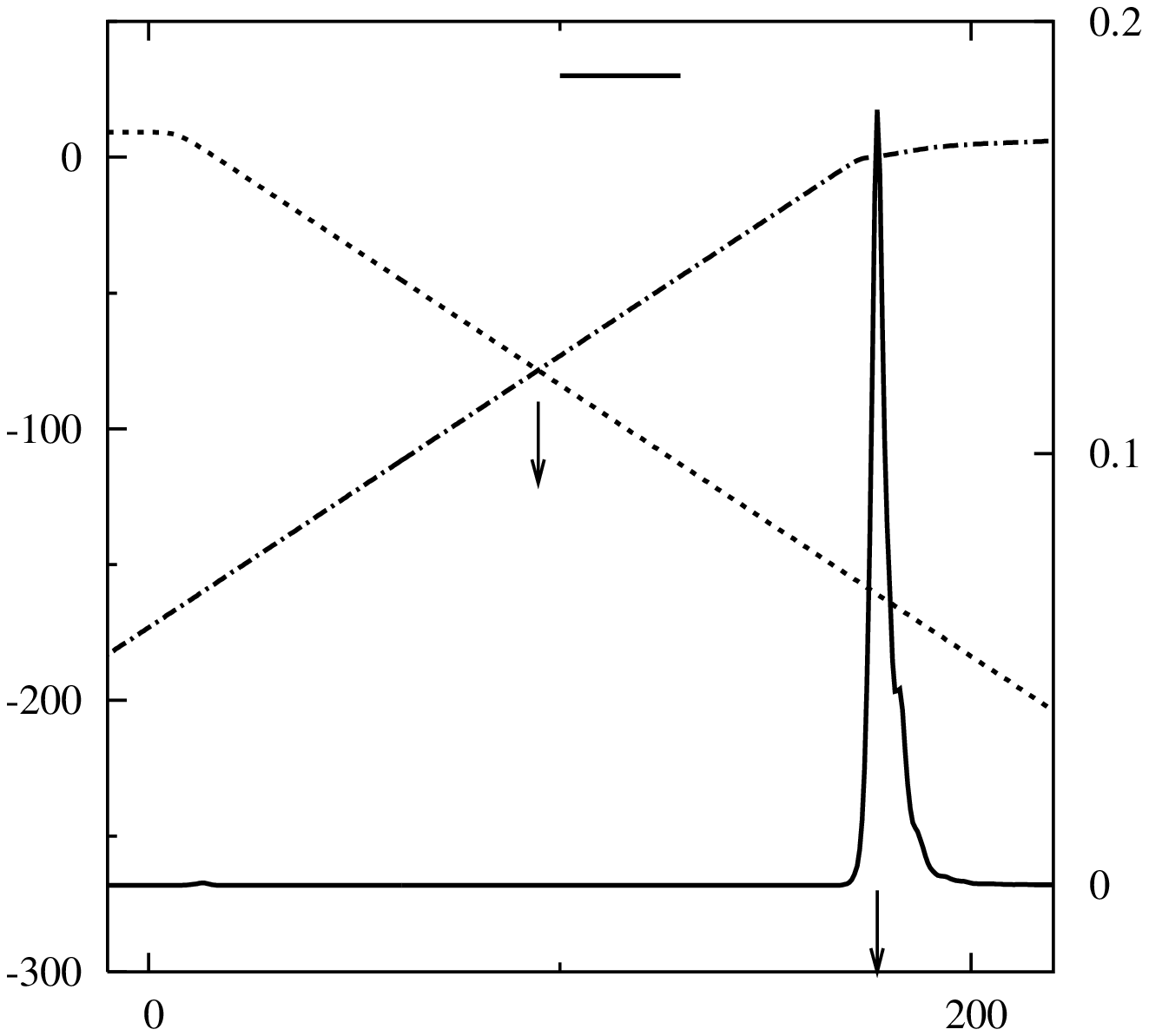,height=7.1cm,width=8.21cm}
    \put(-227,135){\begin{sideways} 
                    $\log \sum$ 
                   \end{sideways}
                   }
    \put(-195,150){\footnotesize $\log \sum_1$}
    \put(-195,105){\footnotesize $\log \sum_0$}
    \put(-165,180){\footnotesize $|{\partial \sigma^2}/{\partial C}|$ }
    \put(-18,145){\large
                   $\left|\frac{\partial \sigma^2}{\partial C}\right|$ 
                   }
    \put(-40,-5){ $C$}
    \put(-128,95){\footnotesize $C_B$}
    \put(-70,5){\footnotesize $C_E$}
  \end{center}
\end{figure}
\clearpage
\begin{figure}
  \begin{center}
    \epsfig{file=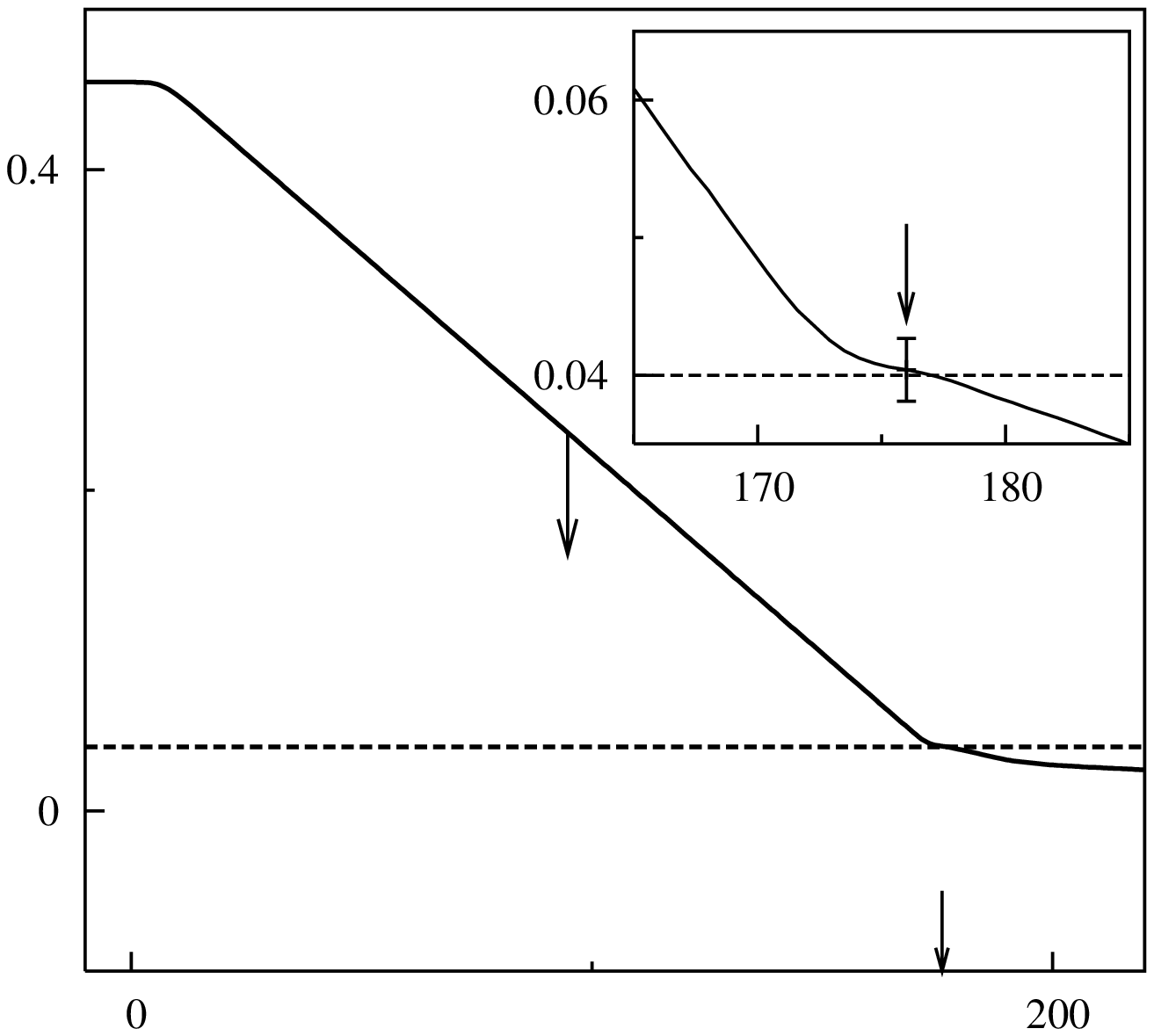,height=7.1cm,width=8.21cm}
    \put(-240,140){\large 
                     $\frac{\Delta F} {A}$ 
                   }
    \put(-120,80){\footnotesize $C_B$}
    \put(-53,35){\footnotesize $C_E$}
    \put(-61,157){\footnotesize $C_E$}
    \put(-38,-7){ $C$}
  \end{center}
\end{figure}
\clearpage
\begin{figure}
  \begin{center}
    \epsfig{file=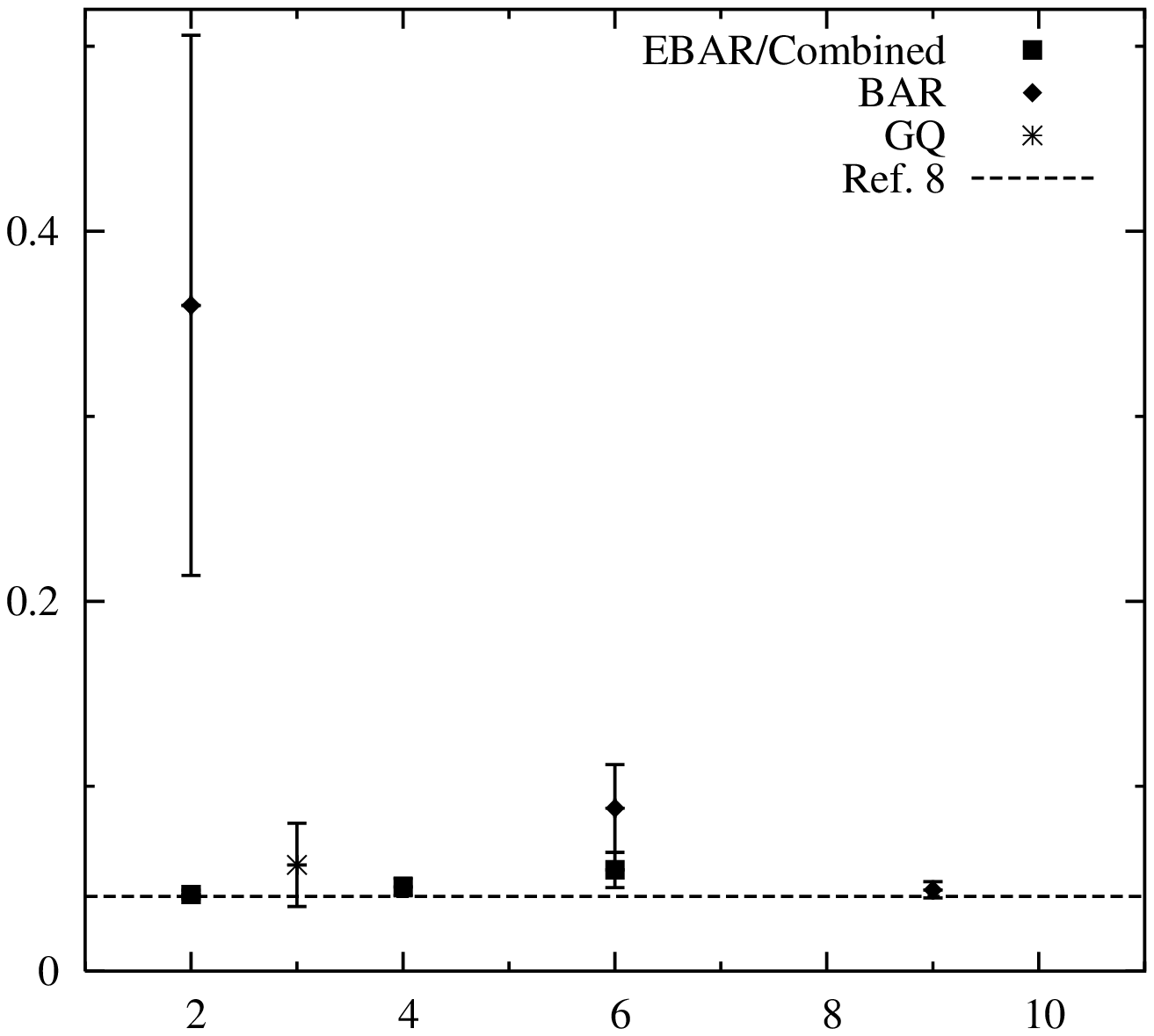,height=7.1cm,width=8.21cm}
    \put(-230,130){\large 
                   $\frac{\Delta F}{A}$ 
                   }
    \put(-60,-7){ $N_s$}
  \end{center}
\end{figure}
\clearpage
\begin{figure}
  \begin{center}
    \epsfig{file=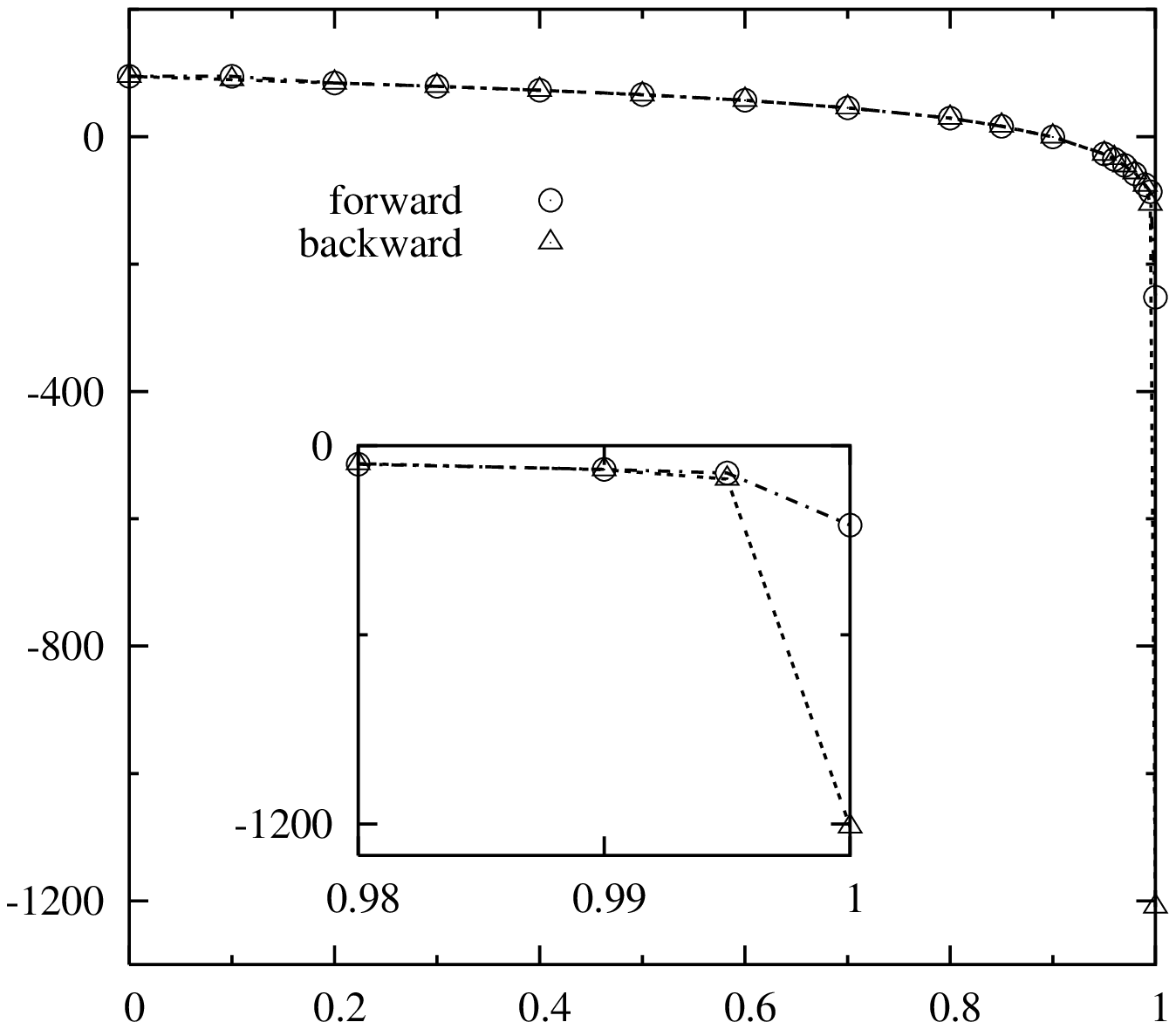,height=7.1cm,width=8.21cm}
    \put(-240,140){\large 
                     $\frac{\partial G^*} {\partial \lambda_3}$ 
                   }
    \put(-38,-7){ {$\lambda_3$}}
  \end{center}
\end{figure}
\clearpage
\begin{figure}
  \begin{center}
    \epsfig{file=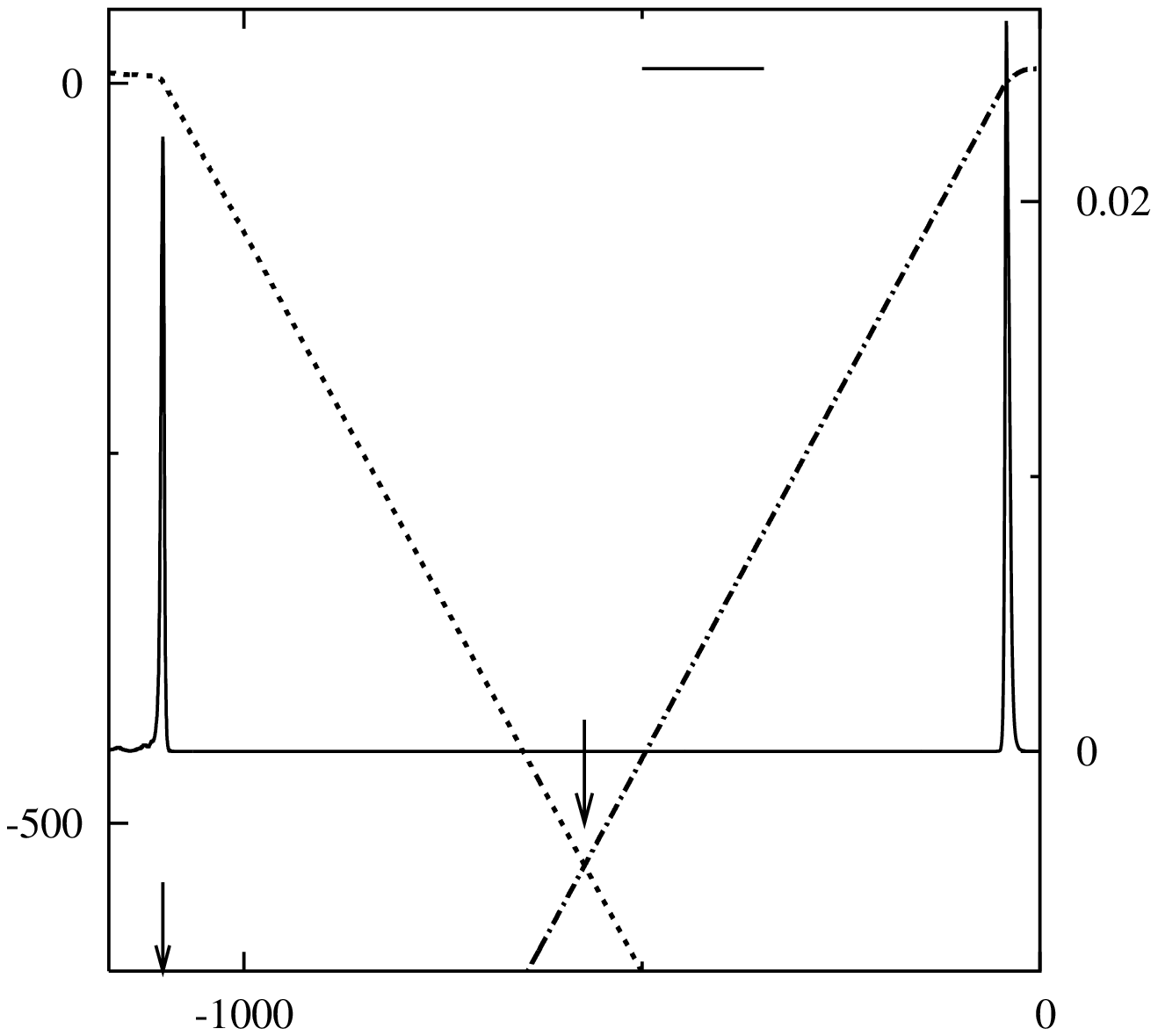,height=7.1cm,width=8.21cm}
    \put(-220,133){\begin{sideways} 
                    $\log \sum$ 
                   \end{sideways}
                   }
    \put(-177,155){\footnotesize $\log \sum_1$}
    \put(-88,155){\footnotesize $\log \sum_0$}
    \put(-150,180){\footnotesize $|{\partial \sigma^2}/{\partial C}|$ }
    \put(-20,133){\large 
                   $\left|\frac{\partial \sigma^2}{\partial C}\right|$ 
                   }
    \put(-120,65){\footnotesize $C_B$}
    \put(-195,35){\footnotesize $C_E$}
    \put(-38,-5){ $C$}
  \end{center}
\end{figure}
\clearpage
\begin{figure}
  \begin{center}
    \epsfig{file=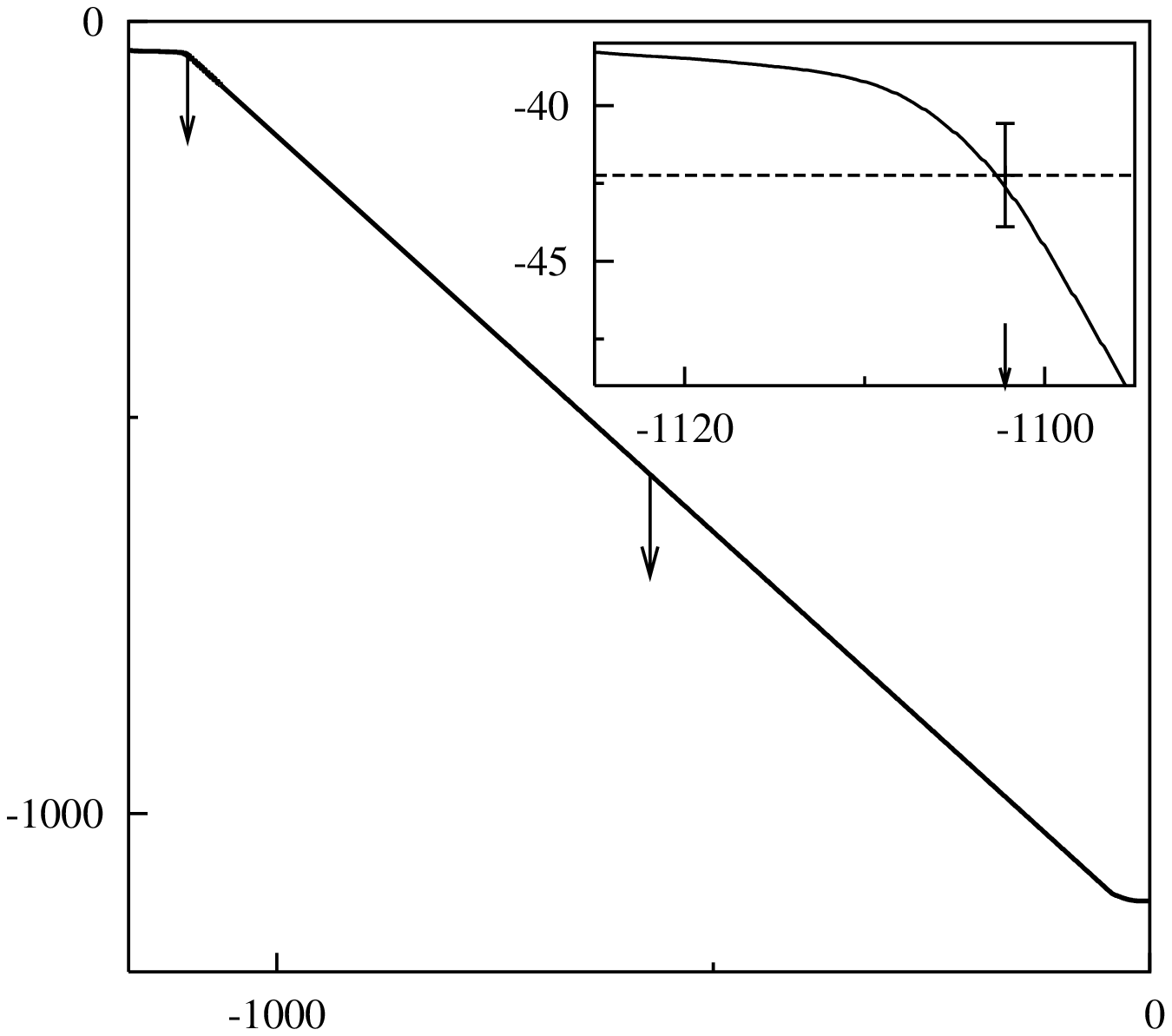,height=7.1cm,width=8.21cm}
    \put(-220,90){\begin{sideways} \small 
                     ${\Delta G} (\lambda_3=0.94 \rightarrow 1)$ 
                   \end{sideways}}
    \put(-108,76){\footnotesize $C_B$}
    \put(-190,157){\footnotesize $C_E$}
    \put(-45,143){\footnotesize $C_E$}
    \put(-38,-7){ $C$}
  \end{center}
\end{figure}
\clearpage
\begin{figure}
  \begin{center}
    \epsfig{file=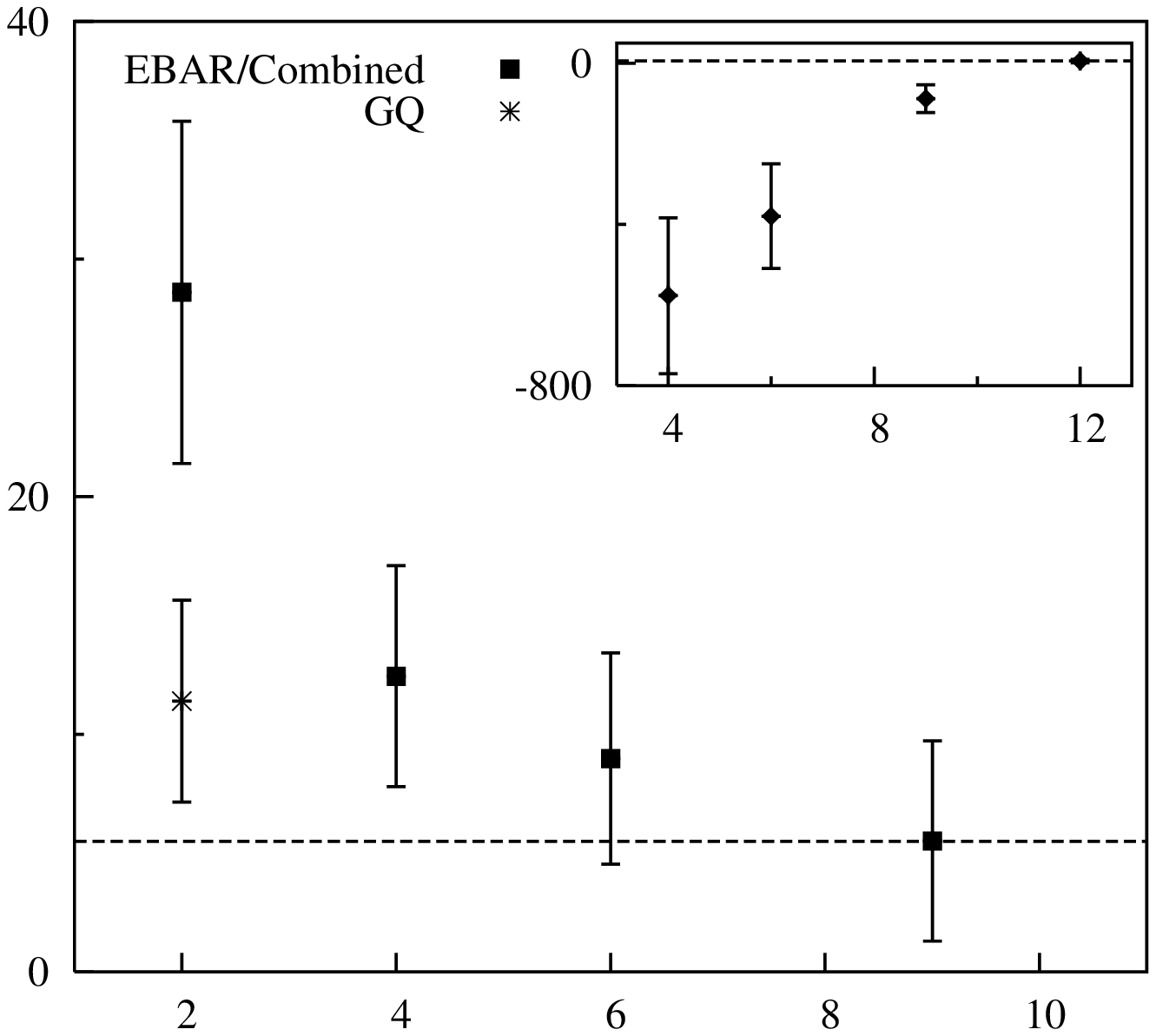,height=7.1cm,width=8.21cm}
    \put(-240,140){\large 
                   ${\Delta G_T}$ 
                   }
    \put(-60,-7){ {$N_s$}}
    \put(-60,150){\footnotesize BAR}
  \end{center}
\end{figure}

\begin{thebibliography}{22}
\expandafter\ifx\csname natexlab\endcsname\relax\def\natexlab#1{#1}\fi
\expandafter\ifx\csname bibnamefont\endcsname\relax
  \def\bibnamefont#1{#1}\fi
\expandafter\ifx\csname bibfnamefont\endcsname\relax
  \def\bibfnamefont#1{#1}\fi
\expandafter\ifx\csname citenamefont\endcsname\relax
  \def\citenamefont#1{#1}\fi
\expandafter\ifx\csname url\endcsname\relax
  \def\url#1{\texttt{#1}}\fi
\expandafter\ifx\csname urlprefix\endcsname\relax\def\urlprefix{URL }\fi
\providecommand{\bibinfo}[2]{#2}
\providecommand{\eprint}[2][]{\url{#2}}

\bibitem[{\citenamefont{Frenkel and Smit}(2002)}]{FRENKEL02}
\bibinfo{author}{\bibfnamefont{D.}~\bibnamefont{Frenkel}} \bibnamefont{and}
  \bibinfo{author}{\bibfnamefont{B.}~\bibnamefont{Smit}},
  \emph{\bibinfo{title}{{Understanding Molecular Simulation}}}
  (\bibinfo{publisher}{Academic Press}, \bibinfo{address}{San Diego},
  \bibinfo{year}{2002}), \bibinfo{edition}{2nd} ed.

\bibitem[{\citenamefont{Fernandez and Harrowell}(2004)}]{FERNANDEZ04}
\bibinfo{author}{\bibfnamefont{J.~R.} \bibnamefont{Fernandez}}
  \bibnamefont{and}
  \bibinfo{author}{\bibfnamefont{P.}~\bibnamefont{Harrowell}},
  \bibinfo{journal}{J. Chem. Phys.} \textbf{\bibinfo{volume}{120}},
  \bibinfo{pages}{9222} (\bibinfo{year}{2004}).

\bibitem[{\citenamefont{Davidchack and Laird}(2003)}]{DAVIDCHACK03}
\bibinfo{author}{\bibfnamefont{R.~L.} \bibnamefont{Davidchack}}
  \bibnamefont{and} \bibinfo{author}{\bibfnamefont{B.~B.} \bibnamefont{Laird}},
  \bibinfo{journal}{J. Chem. Phys.} \textbf{\bibinfo{volume}{118}},
  \bibinfo{pages}{7651} (\bibinfo{year}{2003}).

\bibitem[{\citenamefont{Grochola et~al.}(2005)\citenamefont{Grochola, Snook,
  and Russo}}]{GROCHOLA05}
\bibinfo{author}{\bibfnamefont{G.}~\bibnamefont{Grochola}},
  \bibinfo{author}{\bibfnamefont{I.~K.} \bibnamefont{Snook}}, \bibnamefont{and}
  \bibinfo{author}{\bibfnamefont{S.~P.} \bibnamefont{Russo}},
  \bibinfo{journal}{J. Chem. Phys.} \textbf{\bibinfo{volume}{122}},
  \bibinfo{pages}{064711} (\bibinfo{year}{2005}).

\bibitem[{\citenamefont{Broughton and Gilmer}(1983)}]{BROUGHTON83}
\bibinfo{author}{\bibfnamefont{J.~Q.} \bibnamefont{Broughton}}
  \bibnamefont{and} \bibinfo{author}{\bibfnamefont{G.~H.}
  \bibnamefont{Gilmer}}, \bibinfo{journal}{J. Chem. Phys.}
  \textbf{\bibinfo{volume}{79}}, \bibinfo{pages}{5095} (\bibinfo{year}{1983}).

\bibitem[{\citenamefont{Grochola}(2004)}]{GROCHOLA04}
\bibinfo{author}{\bibfnamefont{G.}~\bibnamefont{Grochola}},
  \bibinfo{journal}{J. Chem. Phys.} \textbf{\bibinfo{volume}{120}},
  \bibinfo{pages}{2122} (\bibinfo{year}{2004}).

\bibitem[{\citenamefont{Apte and Kusaka}(2006)}]{APTE06-1}
\bibinfo{author}{\bibfnamefont{P.~A.} \bibnamefont{Apte}} \bibnamefont{and}
  \bibinfo{author}{\bibfnamefont{I.}~\bibnamefont{Kusaka}},
  \bibinfo{journal}{Phys. Rev. E} \textbf{\bibinfo{volume}{73}},
  \bibinfo{pages}{016704} (\bibinfo{year}{2006}).

\bibitem[{\citenamefont{Apte and Zeng}(2008)}]{APTE08}
\bibinfo{author}{\bibfnamefont{P.~A.} \bibnamefont{Apte}} \bibnamefont{and}
  \bibinfo{author}{\bibfnamefont{X.~C.} \bibnamefont{Zeng}},
  \bibinfo{journal}{Appl. Phys. Lett.} \textbf{\bibinfo{volume}{92}},
  \bibinfo{pages}{221903} (\bibinfo{year}{2008}).

\bibitem[{\citenamefont{Stillinger and Weber}(1985)}]{STILLINGER85}
\bibinfo{author}{\bibfnamefont{F.~H.} \bibnamefont{Stillinger}}
  \bibnamefont{and} \bibinfo{author}{\bibfnamefont{T.~A.} \bibnamefont{Weber}},
  \bibinfo{journal}{Phys. Rev. B} \textbf{\bibinfo{volume}{31}},
  \bibinfo{pages}{5262} (\bibinfo{year}{1985}).

\bibitem[{\citenamefont{Bennett}(1976)}]{BENNETT76}
\bibinfo{author}{\bibfnamefont{C.~H.} \bibnamefont{Bennett}},
  \bibinfo{journal}{J. Comput. Phys.} \textbf{\bibinfo{volume}{22}},
  \bibinfo{pages}{245} (\bibinfo{year}{1976}).

\bibitem[{\citenamefont{Shirts and Pande}(2005)}]{SHIRTS05}
\bibinfo{author}{\bibfnamefont{M.~R.} \bibnamefont{Shirts}} \bibnamefont{and}
  \bibinfo{author}{\bibfnamefont{V.~S.} \bibnamefont{Pande}},
  \bibinfo{journal}{J. Chem. Phys.} \textbf{\bibinfo{volume}{122}},
  \bibinfo{pages}{144107} (\bibinfo{year}{2005}).

\bibitem[{\citenamefont{Mu and Song}(2006)}]{MU06}
\bibinfo{author}{\bibfnamefont{Y.}~\bibnamefont{Mu}} \bibnamefont{and}
  \bibinfo{author}{\bibfnamefont{X.}~\bibnamefont{Song}}, \bibinfo{journal}{J.
  Chem. Phys.} \textbf{\bibinfo{volume}{124}}, \bibinfo{pages}{034712}
  (\bibinfo{year}{2006}).

\bibitem[{\citenamefont{Lu et~al.}(2003)\citenamefont{Lu, Singh, and
  Kofke}}]{LU03}
\bibinfo{author}{\bibfnamefont{N.}~\bibnamefont{Lu}},
  \bibinfo{author}{\bibfnamefont{J.~K.} \bibnamefont{Singh}}, \bibnamefont{and}
  \bibinfo{author}{\bibfnamefont{D.~A.} \bibnamefont{Kofke}},
  \bibinfo{journal}{J. Chem. Phys.} \textbf{\bibinfo{volume}{118}},
  \bibinfo{pages}{2977} (\bibinfo{year}{2003}).

\bibitem[{\citenamefont{Yoo et~al.}(2004)\citenamefont{Yoo, Zeng, and
  Morris}}]{YOO04}
\bibinfo{author}{\bibfnamefont{S.}~\bibnamefont{Yoo}},
  \bibinfo{author}{\bibfnamefont{X.~C.} \bibnamefont{Zeng}}, \bibnamefont{and}
  \bibinfo{author}{\bibfnamefont{J.~R.} \bibnamefont{Morris}},
  \bibinfo{journal}{J. Chem. Phys.} \textbf{\bibinfo{volume}{120}},
  \bibinfo{pages}{1654} (\bibinfo{year}{2004}).

\bibitem[{\citenamefont{Davidchack and Laird}(2000)}]{DAVIDCHACK00}
\bibinfo{author}{\bibfnamefont{R.~L.} \bibnamefont{Davidchack}}
  \bibnamefont{and} \bibinfo{author}{\bibfnamefont{B.~B.} \bibnamefont{Laird}},
  \bibinfo{journal}{Phys. Rev. Lett.} \textbf{\bibinfo{volume}{85}},
  \bibinfo{pages}{4751} (\bibinfo{year}{2000}).

\bibitem[{\citenamefont{Davidchack and Laird}(2005)}]{DAVIDCHACK05}
\bibinfo{author}{\bibfnamefont{R.~L.} \bibnamefont{Davidchack}}
  \bibnamefont{and} \bibinfo{author}{\bibfnamefont{B.~B.} \bibnamefont{Laird}},
  \bibinfo{journal}{J. Phys. Chem. B} \textbf{\bibinfo{volume}{109}},
  \bibinfo{pages}{17802} (\bibinfo{year}{2005}).

\bibitem[{\citenamefont{Handel et~al.}(2008)\citenamefont{Handel, Davidchack,
  Anwar, and Brukhno}}]{HANDEL08}
\bibinfo{author}{\bibfnamefont{R.}~\bibnamefont{Handel}},
  \bibinfo{author}{\bibfnamefont{R.~L.} \bibnamefont{Davidchack}},
  \bibinfo{author}{\bibfnamefont{J.}~\bibnamefont{Anwar}}, \bibnamefont{and}
  \bibinfo{author}{\bibfnamefont{A.}~\bibnamefont{Brukhno}},
  \bibinfo{journal}{Phys. Rev. Lett.} \textbf{\bibinfo{volume}{100}},
  \bibinfo{pages}{036104} (\bibinfo{year}{2008}).

\bibitem[{\citenamefont{Eike et~al.}(2005)\citenamefont{Eike, Brennecke, and
  Maginn}}]{EIKE05}
\bibinfo{author}{\bibfnamefont{D.~M.} \bibnamefont{Eike}},
  \bibinfo{author}{\bibfnamefont{J.~F.} \bibnamefont{Brennecke}},
  \bibnamefont{and} \bibinfo{author}{\bibfnamefont{E.~J.}
  \bibnamefont{Maginn}}, \bibinfo{journal}{J. Chem. Phys.}
  \textbf{\bibinfo{volume}{122}}, \bibinfo{pages}{014115}
  (\bibinfo{year}{2005}).

\bibitem[{\citenamefont{Eike and Maginn}(2006)}]{EIKE06}
\bibinfo{author}{\bibfnamefont{D.~M.} \bibnamefont{Eike}} \bibnamefont{and}
  \bibinfo{author}{\bibfnamefont{E.~J.} \bibnamefont{Maginn}},
  \bibinfo{journal}{J. Chem. Phys.} \textbf{\bibinfo{volume}{124}},
  \bibinfo{pages}{154504} (\bibinfo{year}{2006}).

\bibitem[{\citenamefont{Apte and Kusaka}(2005)}]{APTE05}
\bibinfo{author}{\bibfnamefont{P.~A.} \bibnamefont{Apte}} \bibnamefont{and}
  \bibinfo{author}{\bibfnamefont{I.}~\bibnamefont{Kusaka}},
  \bibinfo{journal}{J. Chem. Phys.} \textbf{\bibinfo{volume}{123}},
  \bibinfo{pages}{194503} (\bibinfo{year}{2005}).

\bibitem[{\citenamefont{Broughton and Li}(1987)}]{BROUGHTON87}
\bibinfo{author}{\bibfnamefont{J.~Q.} \bibnamefont{Broughton}}
  \bibnamefont{and} \bibinfo{author}{\bibfnamefont{X.~P.} \bibnamefont{Li}},
  \bibinfo{journal}{Phys. Rev. B} \textbf{\bibinfo{volume}{35}},
  \bibinfo{pages}{9120} (\bibinfo{year}{1987}).

\bibitem[{\citenamefont{Press et~al.}(1992)\citenamefont{Press, Teukolsky,
  Vetterling, and Flannery}}]{PRESS92}
\bibinfo{author}{\bibfnamefont{W.~H.} \bibnamefont{Press}},
  \bibinfo{author}{\bibfnamefont{S.~A.} \bibnamefont{Teukolsky}},
  \bibinfo{author}{\bibfnamefont{W.~T.} \bibnamefont{Vetterling}},
  \bibnamefont{and} \bibinfo{author}{\bibfnamefont{B.~P.}
  \bibnamefont{Flannery}}, \emph{\bibinfo{title}{{Numerical Recepies}}}
  (\bibinfo{publisher}{Cambridge University Press},
  \bibinfo{address}{Cambridge}, \bibinfo{year}{1992}).

\end{thebibliography}
\end{document}